\documentclass[10pt,aps,physrev,raggedbottom,longbibliography,nobalancelastpage,reprint,citeautoscript,letterpaper,superscriptaddress]{revtex4-2}

\usepackage{multirow}
\usepackage{bbold}
\usepackage{graphicx}
\usepackage{subfigure}

\usepackage[usenames,dvipsnames]{color}
\usepackage{graphicx,microtype}
\usepackage{amsmath}
\usepackage[bookmarks=false,colorlinks]{hyperref}
\usepackage{lmodern}
\usepackage[all]{hypcap} 
\hypersetup{linkcolor=magenta,citecolor=MidnightBlue,filecolor=Plum,urlcolor=MidnightBlue}

\usepackage[scaled=0.875]{helvet}


\usepackage{ccaption}
\usepackage{ragged2e}
  \captionnamefont{\small \sffamily \bfseries }
  \captiontitlefont{\small }
  \captiondelim{~\textbar~}
  \captionstyle{\justifying}

\begin{document}
\preprint{APS/123-QED}

\title{Generation of magnetic metal-organic frameworks}

\author{Alexander C.\ Tyner}
\affiliation{NORDITA, KTH Royal Institute of Technology and Stockholm University, Stockholm, Sweden}
\author{Avinash Pathapati}
\affiliation{Doublet labs, Stockholm, Sweden}
\author{Alexander. V. Balatsky}
\affiliation{NORDITA, KTH Royal Institute of Technology and Stockholm University, Stockholm, Sweden}


\date{\today}

\begin{abstract} 
The potential to utilize metal-organic frameworks as a replacement for rare earth materials as well as in technological applications has prompted increased interested in this material class. The simulation of organic materials, including metal-organic frameworks (MOFs), represents a computational challenge due to an increased average number of atoms in the unit cell. Compounding this challenge, modern materials databases are generally limited to inorganic structures due to their utility in modern technologies such as batteries and integrated circuits. Machine-learning tools appear ideally suited to study these systems. However, organic materials are generally underrepresented in the training sets of foundational models. In this work we leverage the the Organic Materials Database (OMDB) to create a training dataset comprised of more than 15,000 single-point first-principles computations for finetuning machine learned interatomic potentials. Specifically, we fine tune CHGNet and implement a site substitution workflow to identify novel, highly magnetic, MOFs from structural prototypes within the QMOF database.  
\end{abstract}

\maketitle

\section{Introduction} 
Metal-organic frameworks (MOFs) have emerged as a material class of supreme interest from both an academic and industrial perspective. MOFs represent polymers containing metallic clusters connected by organic ligands\cite{kitagawa2014metal,furukawa2013chemistry,czaja2009industrial,zhou2012introduction,schneemann2014flexible,kuppler2009potential,mueller2006metal,cui2016metal,james2003metal,kreno2012metal,lee2009metal,long2009pervasive,wang2009postsynthetic,kurmoo2009magnetic,cohen2012postsynthetic,wang2017metal,corma2010engineering,dang2017nanomaterials,jiao2018metal,yuan2018stable}. This interesting intersection of organic and inorganic chemistry gives rise to novel thermodynamic and electronic properties. A primary goal for MOF design is the achievement of ferromagnetism rivaling rare-earth compounds\cite{kurmoo2009magnetic,espallargas2018magnetic,thorarinsdottir2020metal}. Such a material would offer a lower-cost and greener alternative to a fundamental component of modern technologies.  
\par 
Design and discovery of MOFs, particularly for electronics applications, is extremely challenging due to the size of the primitive unit cell, often containing more than 100 atoms. The computational expense of first-principles simulations of electronic structure, namely density functional theory, scale cubically, $O(N^3)$, in atom number, limiting high-throughput screenings. Machine learning tools, particularly machine learned potentials (MLPs)\cite{chen2022universal,deng2023chgnet,batatia2022mace,chen2019graph,batatia2025foundation}, appear ideally suited to enable MOF design\cite{rosen2022high,rosen2021machine}. However, organic materials in general, and MOFS more specifically, are underrepresented in the datasets used to train available foundational MLPs such as MACE\cite{batatia2022mace}, CHGNet\cite{deng2023chgnet}, M3GNet\cite{chen2022universal}, and Equiformer\cite{liao2022equiformer}. 
\par 
In this work we present a new dataset, OMDB-Traj, for fine-tuning foundational MLPs. This dataset leverages the power of the Organic Materials Database (OMDB)\cite{borysov2017organic,omdb5,omdb4,omdb2,omdb1,TynerSC}, one of the largest public databases of organic materials containing more than 40,000 entries. While past work has performed data-mining on the OMDB for purposed of training property prediction machine-learning networks, this work represents the first use of the OMDB for training a MLP, namely CHGNet (v0.3.0).
\par 
We choose CHGNet as it is unique in its incorporation of magnetic moments, data included in the OMDB-Traj training set. After demonstrating that fine-tuning  dramatically increases performance of the MLP on a test set of organic compounds, we implement the training model in a generative workflow. This workflow takes as input all magnetic compounds within the QMOF database. Each compound is then treated as a prototype for which substitution of inequivalent atomic sites hosting magnetic atoms for alternative magnetic atoms is performed. Upon substitution the structure is relaxed using CHGNet and the magnetic moments of the relaxed structure are predicted. Multiple permutations are examined for each prototype, resulting in more than 149K novel structures. From this set we collect the 100 displaying the highest magnetic moment and perform validation of stability and DFT verification of the magnetic moments.

\begin{figure*}
    \centering
    \includegraphics[width=16cm]{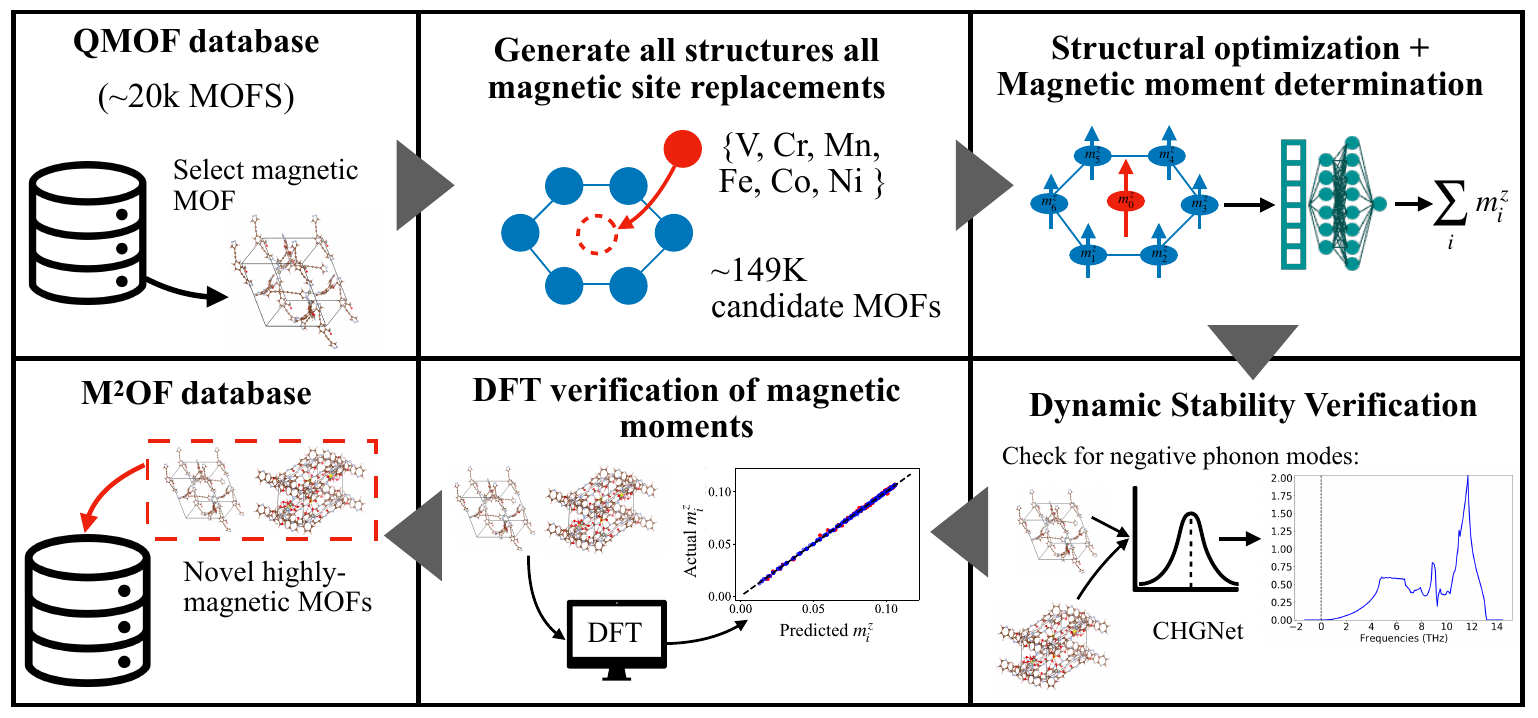}
    \caption{Workflow for generation of novel magnetic MOFs through site substitution of magnetic prototypes within the QMOF database. The finetuned machine learned interatomic potential, CHGNet, is leveraged for determination of magnetic moments and isolation of dynamically stable structures, reducing computational expense in the final density functional theory computations.   }
    \label{fig:workflow}
\end{figure*}

\section{Methodology}

\subsection{Training Dataset}
\begin{figure*}
    \centering
    \includegraphics[width=16cm]{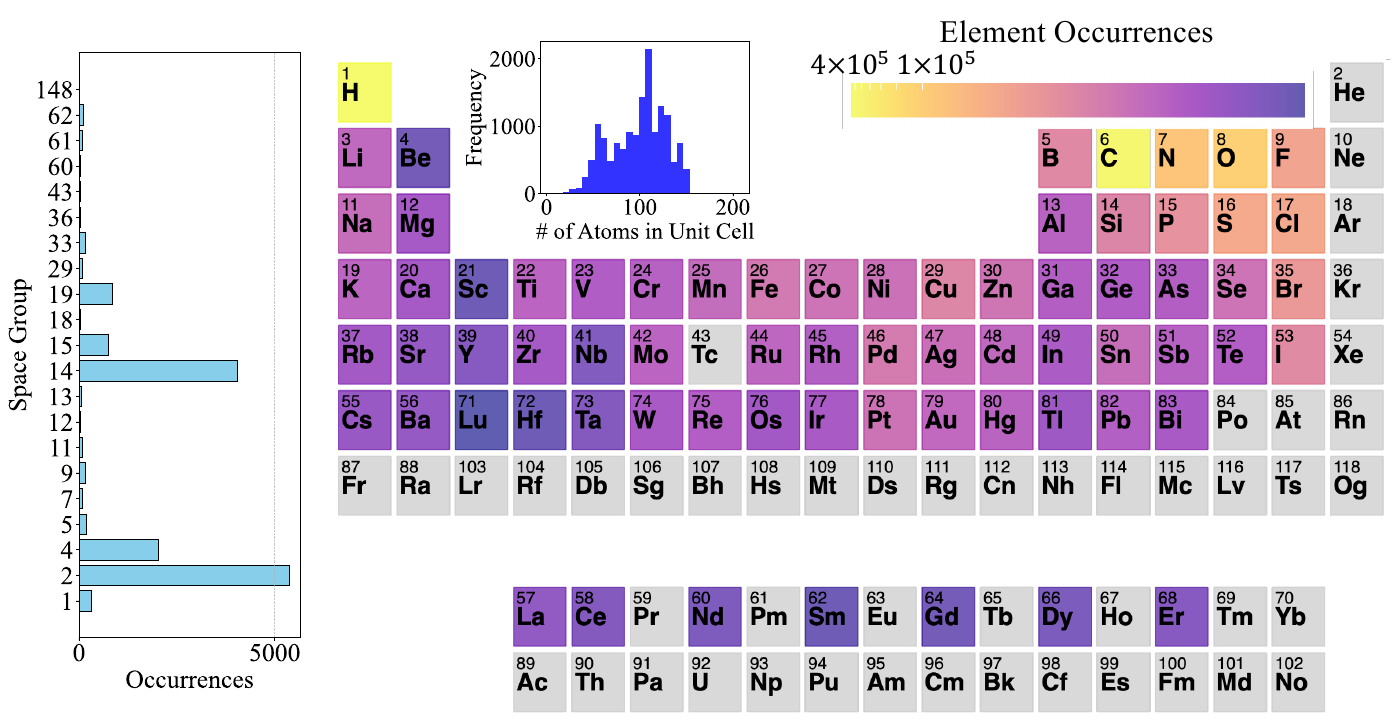}
    \caption{Statistical makeup of the OMDBTraj dataset including spacegroups, atoms in the primitive unit cell and element frequency. The increased frequency of organic elements, large numbers of atoms in the primitive cell and low-symmetry space groups is representative of organic systems.}
    \label{fig:dataset}
\end{figure*}

The data used to fine-tune CHGNet V0.3.0 in this work consists of 15,000 single point DFT computations from the OMDB. These density functional computations are performed using the Vienna Ab initio simulation package (VASP)\cite{hafner2008ab}. The exchange-correlation functional was approximated by the generalized gradient approximation (GGA) according to Perdew, Burke and Ernzerhof\cite{Perdew1996} and a $ 6 \times \times 6 \times 6$ $\Gamma$-centered Monkhorst-Pack grid\cite{monkhorst1976special} is used for the self-consistent cycle.  Statistics of the dataset are shown in Fig. \eqref{fig:dataset}, demonstrating wide coverage of the periodic table and diversity in both spacegroup and unit-cell size. These 15,000 data-points are selected from the database as they contain information regarding the magnetic moments and exist near equilibrium with forces all less than 20 eV/$\AA$.
\subsection{Model Performance}
\begin{figure}
    \centering
    \includegraphics[width=8cm]{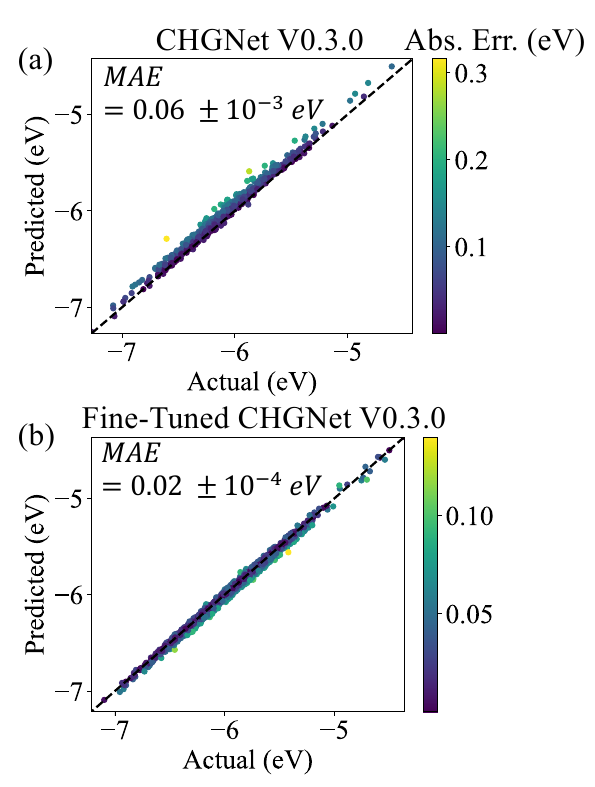}
    \caption{(a) Comparison of CHGNet V0.3.0 energies with density functional theory for test set of 1500 structures taken from OMDBTraj. (b) Comparison of CHGNet V0.3.0 after finetuning on OMDBTraj with the same test set.  }
    \label{fig:ModelPerformance}
\end{figure}

The model is finetuned for 50 epochs using a cosine annealing learning rate with a maximum value of $1\times10^{-3}$ and a mean-average error loss function. We choose to increase the contribution of the mean-average error from the forces by a factor of 100 relative to that from energy, stress, or magnetic moments as forces are generally a large source of error when fine-tuning from a foundational model trained using mostly inorganic crystalline systems due to the flexible nature of MOFs. The final results of the fine-tuning are shown in Tab. \eqref{tab:opt} and Fig. \eqref{fig:ModelPerformance}.

\subsection{Material Discovery}
Inverse design of crystalline systems has progressed rapidly in recent years with the introduction of diffusion models\cite{xie2021crystal,zeni2025generative}, LLM based models\cite{gruver2024fine}, adversarial networks\cite{zhao2021high,Tyner_2026} and more\cite{merchant2023scaling}. While such deep generative approaches are remarkable in their ability to generate highly novel crystals, the generated compounds often lack symmetry and are unstable or difficult to synthesize. These issues are compounded when dealing with large atomic structures and has led to limited application of such methods to MOFs. This is our motivation for pursuing a low-tech but highly effective alternative strategy, site-replacement. Rather than generating completely novel crystal from a latent space, we begin with a prototype structure, specifically a magnetic MOF from the QMOF database. Next, all equivalent atomic sites hosting V, Co, Ni, Fe, or Mn atoms are identified. New structures are then generated considering all possible permutations in which magnetic elements at equivalent sites are substituted for an element from this list. As MOFs are generally large structures with little symmetry, this produces $> 149$K total structures. 
\par 
For each of the resulting structures, we optimize the atomic sites and lattice parameters using the finetuned MLP until all forces are below $0.1 eV/\AA$ and predict the magnetic moments of the final structure. The 100 systems with the largest total magnetic moment are then identified and isolated for validation within density functional theory as well as analysis of dynamic stability. A schematic detailing this workflow is given in Fig. \eqref{fig:workflow}.
\begin{table}
\caption{ \textbf{Finetuning the CHGNET MLiP.} Results of finetuning CHGNET using the OMDBTraj dataset.}
\begin{ruledtabular}
\begin{tabular}{lll}
   & Fine Tuned MAE & Base MAE \\ 
\hline
Energies & 0.02 eV        &  0.06 eV\\
Forces    & 0.081 ev/\AA        &  0.097ev/\AA \\
Magnetic Moments   & 0.001 $\mu$ B         &   0.010 $\mu$ B  \\ 

\end{tabular}
\end{ruledtabular}
\label{tab:opt}
\end{table}

\section{Material Results}
\begin{figure}
    \centering
    \includegraphics[width=8cm]{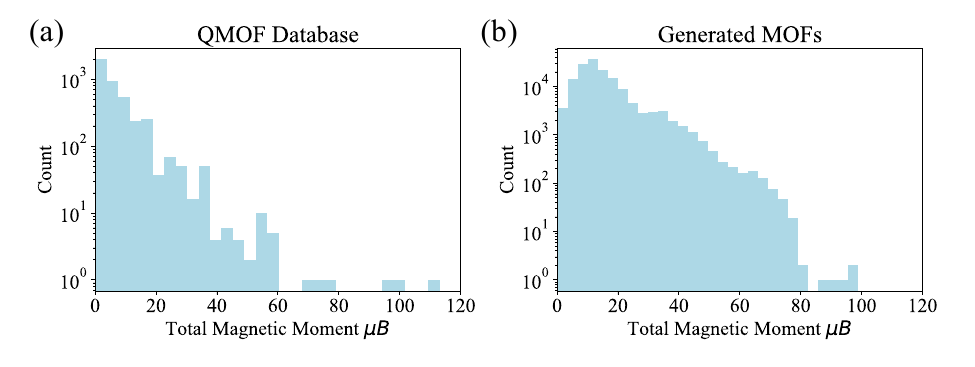}
    \caption{(a) Distribution of magnetic moment amplitude sum for structures within the QMOF database. (b) Distribution of magnetic moment amplitude sum for novel structures formed from magnetic element substitution with QMOF database.}
    \label{fig:Histogram}
\end{figure}

From the 149,420 materials generated, we find the distribution of total magnetic moments given in Fig. \eqref{fig:Histogram}. We select the 100 systems with the largest total magnetic moment and validate the dynamic stability of these structures by computing the phonon density of states using Phonopy\cite{togo2023first} and the fine-tuned CHGNet model. Following Ref. \cite{elena2025machine}, for systems admitting negative phonon modes, we perform a low-temperature NVE ensemble molecular dynamics simulation at 7.5K using the finetuned CHGNet and re-optimize the structure before recomputing the phonon modes. If this process fails to eliminate negative phonon modes after three iterations the structure is eliminated. This process reduces the number of compounds under investigation to 16.

\begin{figure*}
    \centering
    \includegraphics[width=18cm]{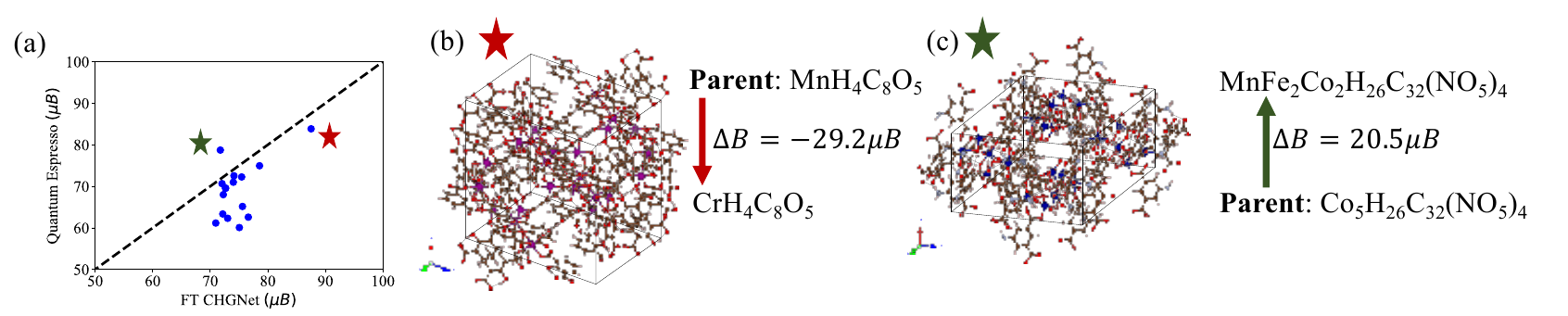}
    \caption{(a) Comparison of total magnetic moment predicted by the finetuned CHGNet machine learned potential (MLP) and by a DFT computation using Quantum Espresso for the 16 compounds deemed dynamically stable through computation of the phonon modes using the MLP. (b) Compound supporting largest magnetic moment in (a), marked by a red star. Despite hosting a large magnetic moment, this value is reduced from that of the parent system in the QMOF database. (c) Compound supporting second largest magnetic moment in (a), marked by a green star. The low-symmetry of this compound allows for a complex site substitution mapping that successfully elevates the total magnetic moment above that of the parent system in the QMOF database.   }
    \label{fig:Validation}
\end{figure*}
\par 
From the 16 compounds deemed dynamically stable by the finetuned CHGNet, we directly compute the magnetic moments with density functional theory (DFT) computations via the Quantum Espresso software package\cite{QE-2020}. For details of the DFT computations please consult the methods. The results of this validation study are shown in Fig. \eqref{fig:Validation}(a). The compounds displaying the two highest magnetic moments, marked by red and green stars in Fig. \eqref{fig:Validation}(a), are shown in Fig.\eqref{fig:Validation}(b)-(c) respectively. We further provide the structural formula of the parent system within the QMOF database and change in total magnetic moment from the parent system to the compound generated by site substitution. We note that not all site substitutions have the effect of increasing the total magnetic moment, an example of this is Fig.\eqref{fig:Validation}(b). Nevertheless, site substitution can provide alternative benefits, such as enhanced stability or synthesizability while maintaining a large total magnetic moment.  By contrast, Fig. \eqref{fig:Validation}(c) shows how the low symmetry of MOFs can lead to a large diversity of possible site substitutions with the potential to produce a larger magnetic moment.  
\section{Summary}
In this work we have leveraged the power of the data contained within the OMDB for fine-tuning a state-of-the-art MLP. The utility of such a machine-learning tool in the context of MOFs is then demonstrated by pursuing a generative workflow centered around discovery of highly magnetic systems. In this process we have identified and structurally optimized 149,420 structures, all of which are contained within the $M^{2}MOF$ database, available within the OMDB, along with the predicted magnetic moments. We have then validated the results of the most promising compounds via first-principles computations. We expect this workflow to applicable to a range of other properties given the utility and growing interest in MOFs for technological applications.

\begin{acknowledgments}

NORDITA is supported in part by NordForsk. The computations were enabled by resources provided by the National Academic Infrastructure for Supercomputing in Sweden (NAISS), partially funded by the Swedish Research Council through grant agreement no. 2022-06725.

\end{acknowledgments}

\section{Data availability}
A subset of the generated MOFs is available at Ref.~\cite{doublet}. Portions of the data utilized in this work are commercially sensitive but available upon request from the authors.

\appendix
\section{Methods}
Density functional theory (DFT) computations to validate the magnetic moments of MOFs deemed dynamically stable by the fine-tuned machine learned potential are performed using the Quantum Espresso software package \cite{QE-2020,Perdew1996}. The exchange-correlation functional was approximated by the generalized gradient approximation (GGA) according to Perdew, Burke and Ernzerhof\cite{Perdew1996} and a $ 2 \times \times 2 \times 2$ $\Gamma$-centered Monkhorst-Pack is used. 
Ultra-soft pseudopotentials from Ref. \cite{garrity2014pseudopotentials} are employed along with a plane-wave energy cutoff of 40\,Ry.

\bibliography{Ref.bib}

\end{document}